\documentclass[pra,12pt,floatfix]{revtex4}
\usepackage{graphicx}
\usepackage{amsmath}

\begin{document}
\title{Evolution of electronic and ionic structure of $Mg$-clusters with
the growth cluster size}

\author {Andrey Lyalin}
\altaffiliation[Permanent address: ]
{Institute of Physics, St Petersburg State University,
198504 St Petersburg, Petrodvorez, Russia}
\email[Email address: ]{lyalin@th.physik.uni-frankfurt.de}

\author {Ilia A Solov'yov}
\altaffiliation[Permanent address: ]
{A. F. Ioffe Physical-Technical Institute, 194021 St. Petersburg, Russia}
\email[Email address: ]{ilia@th.physik.uni-frankfurt.de}

\author {Andrey V Solov'yov}
\altaffiliation[Permanent address: ]
{A. F. Ioffe Physical-Technical Institute, 194021 St. Petersburg, Russia}
\email[Email address: ]{solovyov@th.physik.uni-frankfurt.de}

\author {Walter Greiner}
\affiliation{Institut f\"{u}r Theoretische Physik der Universit\"{a}t
Frankfurt am Main, Robert-Mayer Str. 8-10, D-60054 Frankfurt am Main, Germany}

\begin{abstract}
The optimized structure and electronic properties of neutral and singly charged
magnesium  clusters have been investigated
using {\it ab initio} theoretical methods based on
density-functional theory and systematic post-Hartree-Fock many-body perturbation theory
accounting for {\it all} electrons in the system.
We have systematically calculated the optimized geometries of neutral
and singly charged magnesium clusters consisting of up to 21 atoms,  electronic 
shell closures, binding energies per atom, ionization potentials and the gap 
between the highest occupied and the lowest unoccupied molecular orbitals.
We have investigated the transition to the {\it hcp} structure and 
metallic evolution of the magnesium clusters, as well as the stability
of linear chains and rings of magnesium atoms.
The results obtained are compared with the available experimental data and 
the results of other theoretical works.
\end{abstract}

\pacs{36.40.Cg, 36.40.Mr, 31.10.+z, 31.15.Ne}

\keywords{local-density approximation; many-body theory; metal clusters}

\maketitle

\section{Introduction}
\label{intro}

Metal clusters have been recognized as new
physical objects with their own properties almost two decades ago.
This became clear after such experimental successes as
the discovery of electronic shell structure in metal clusters
\cite{Knight84}, observation of plasmon resonances
\cite{Br89,Selby89,Selby91}, formation of singly and doubly charged 
negative cluster ions \cite{Lutz} and many more.
Comprehensive survey of the field can be found 
in review papers and books; see, e.g.,
\cite{deHeer93,Brack93,BrCo94,Haberland94,Guet97,MetCl99,LesHouches,Jellinek99,Meiwes-Broer00}.


The electronic shell structure of metal clusters has been discovered in \cite{Knight84}
by the observation of the strong pics in the mass spectra of sodium clusters.
The enhanced stability of some clusters, the so-called magic clusters, was 
explained by the closure of shells of delocalized electrons.
A simple physical model describing electronic shell structure of 
metal clusters has been developed within the jellium approximation
(see, e.g., \cite{MetCl99}) by analogy with 
the shell model of atomic nuclei (see, e.g., \cite{Eisenberg_Greiner}). 
The jellium model is very successful for the simple alkali 
metals ($Na$, $K$), for which one electron per 
atom is delocalized \cite{LSSCG00,Lyalin01a,MLSSG02}. 
The jellium model electronic shell closures for 
alkali-metal clusters define the magic numbers $N=$ 8, 20, 34, 40, 58, 92 
that are in a good agreement with experiment. Note that the jellium model can be 
generalized by accounting for the collective ion background vibration dynamics 
\cite{GSG99,GISG00} and be used as a very appropriate framework for the investigating
collision processes involving atomic clusters \cite{AVSol}.

Clusters of divalent metals are expected to differ from the 
jellium model predictions at least at small cluster sizes. In this case,  
bonding between atoms is expected to have some features of the
van der Waals type of bonding, because the electronic shells in the divalent atoms are filled. 
Thus, clusters of divalent metals are very appropriate
for studing non-metal to metal transition,
testing different theoretical methodologies and conceptual developments of atomic
cluster physics. However, relatively little work was done so far on the exploration of the 
alkali-earth metal clusters in comparison with that for the alkali-metal clusters; see, e.g.,
\cite{MetCl99,StructNa} and references therein.

Previous theoretical studies of the magnesium cluster properties have been performed 
using pseudopotential approximation for the treatment of inner electrons in a cluster and  
the density-functional theory for the description of outer shell electrons.
The electronic properties, geometry and stability of small $Mg$ metal clusters with the
number of atoms $N \le 7$
have been investigated in \cite{Reuse89,Reuse90} using the pseudopotential 
local-spin-density approximation. 
The geometrical structure and bonding nature of $Mg_N$ clusters with 
$N$ up to 13 have been studied in \cite{Kumar91} 
using the density-functional molecular-dynamics method. The size evolution of 
bonding in magnesium clusters $Mg_N$ with $N=8-13, 16, 20$ have been 
studied in \cite{Delaly92} using the local-density approximation that accounts for gradient
corrections. Structural and electronic properties of small magnesium clusters
(${N \le 13}$) were studied in \cite{Akola01} using a first-principles simulation method 
in conjunction with the density-functional theory and the generalized gradient correction
approximation for the exchange-correlation functional. It was shown \cite{Akola01} 
that the metallization in magnesium clusters has a slow and nonmonotonic evolution,
although, also jellium-type magic clusters were observed \cite{Kumar91,Delaly92}. 
In order to extend such calculations to larger systems, symmetry restricted methods 
have been developed. The spherically-averaged-pseudo-potential scheme with the local and 
non-local pseudopotentials has been used for the investigation of the electronic structure and 
shell closures of spherical $Mg_N$ clusters up to $N=46$ \cite{Serra02}.
  
Recently, new experimental data for $Mg$ clusters have been obtained, 
indicating the most intensive pics in the mass 
spectra at $N=$ 5, 10, 15, 18, 20, 25, 28, 30, 35, 40, 47, 56, 59, 62, and 69 
\cite{Diederich01}.
These numbers deviate from the sequence of magic numbers which were
obtained for simple alkali metal
clusters, and cannot be reproduced within simple jellium models.
This fact was interpreted in \cite{Diederich01,Doppner01}  
within the spherical shell model by diving of the high angular momentum states down
through the states with lower $l$.

In the present work we investigate the optimized ionic structure
and the electronic properties of neutral and singly charged magnesium 
clusters  within the size range $N \le 21 $.
We calculate  binding energies per atom, ionization potentials and energy gaps 
between the highest occupied and the lowest unoccupied molecular orbitals.
Our calculations are based on {\it ab initio} theoretical methods invoking the
density-functional theory and systematic post-Hartree-Fock many-body theory
accounting for {\it all} electrons in the system.
The results obtained are compared
with the available experimental data and the results of other theoretical works.

The atomic system of units, $|e|=m_e=\hbar=1$, has been used
throughout the paper, unless other units are indicated.

\section{Theoretical methods}
\label{theory}

Our calculations have been 
performed  with the use of
the Gaussian 98 software package \cite{Gaussian98}.   
We have utilized the $6-311G(d)$ basis set of 
primitive Gaussian functions to expand the cluster orbitals 
\cite{Gaussian98,Gaussian98_man}.

The cluster geometries have been determined 
by finding local minima on the multidimensional
potential energy surface for a cluster. 
We have taken into account {\it all electrons} available in the system, 
when computing the potential energy surface.
With increasing cluster size, such calculations become computer
time demanding. In this work, we limit the calculations by the cluster size 
$N = 21$.

The key point of calculations is fixing the starting geometry of the cluster, 
which could converge during the calculation to a local or the global minimum.
There is no unique way for achieving this goal with {\it Gaussian 98} 
\cite{Gaussian98_man}.
In our calculations, we have created the starting geometries empirically,
often assuming certain cluster symmetries. Note, that during
the optimization process the geometry of the cluster as well
as its initial symmetry sometimes change dramatically.
All the characteristics of clusters, which we have calculated
and present in next section, are obtained for the clusters with optimized 
geometry.

In this work we concentrate on the systematic exploration 
of the properties of magnesium clusters using the density-functional theory
based on the hybrid Becke-type three-parameter exchange functional
\cite{Becke88} paired with the gradient-corrected Lee, Yang and Parr correlation
functional ($B3LYP$) \cite{LYP,Parr-book}, as well as the gradient-corrected 
Perdew-Wang 91 correlation functional ($B3PW91$)  \cite{PW91,PerWan}.
The important feature of the density-functional method consists in the fact 
that it takes into account many-electron correlations via the
phenomenological exchange-correlation potential.
However, so far, there has not been found the unique potential,
universally applicable for different systems and conditions.
As a result there are many different parameterizations for the exchange-correlation
potential valid for special cases. 

Alternatively, we use a direct {\it ab initio} method
for the description of electronic properties of metal clusters,
which is based on the consistent post-Hartree-Fock 
many-body theory \cite{MP}. In the present work, we apply the {M\o ller-Plesset} 
perturbation theory of the fourth order ($MP4$). 
Based on the fundamental physical principles 
being free from any phenomenological parameters, 
this model can be refined by extending the quality of the
approximations, while the physical meaning of the effects included is
clearly demonstrated. Thus, often such an approach predicts more accurate and reliable 
characteristics of metal clusters than the density-functional theory. 

In the present work we use both different theoretical schemes for calculations 
taking advantage of the clear physical meaning and reliability of 
the post-Hartree-Fock perturbation theory and the numerical efficiency
of the density-functional methods.

\section{Numerical results and discussion}
\label{results}

\subsection{Geometry optimization of Mg$_{N}$ and Mg$_{N}^{+}$ clusters}
\label{geom_opt}

\begin{figure}[h]
\includegraphics[scale=0.4]{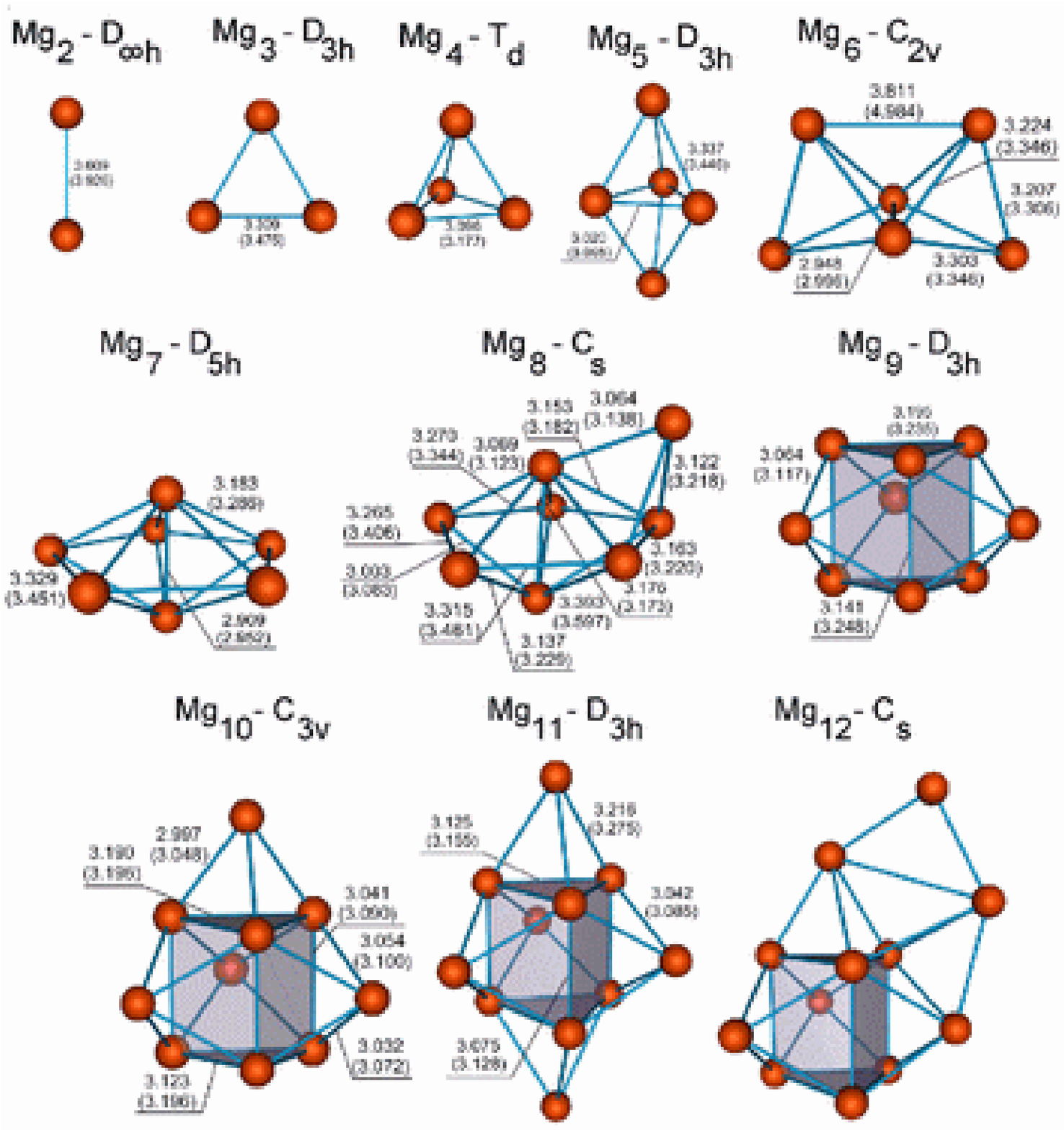}
\includegraphics[scale=0.4]{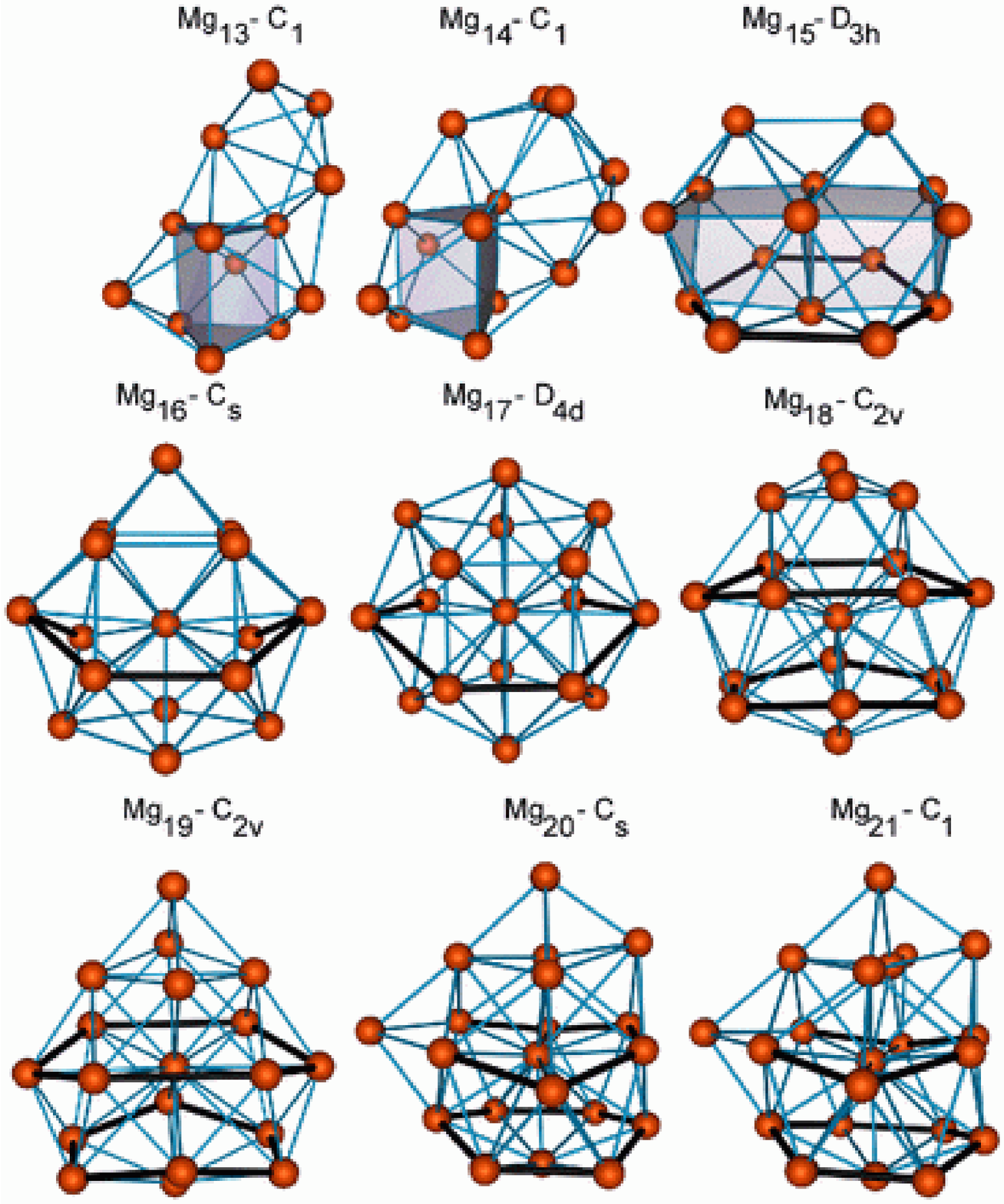}
\caption{Optimized geometries of the neutral magnesium clusters
$Mg_{2} - Mg_{21}$ calculated 
in the $B3PW91$ approximation. The interatomic distances are given in angstroms.
The values in brackets correspond to the $B3LYP$ results.
The label above each cluster image indicates the point symmetry group of the cluster.}
\label{geom_neutral}
\end{figure}

\begin{figure}[h]
\includegraphics[scale=0.4]{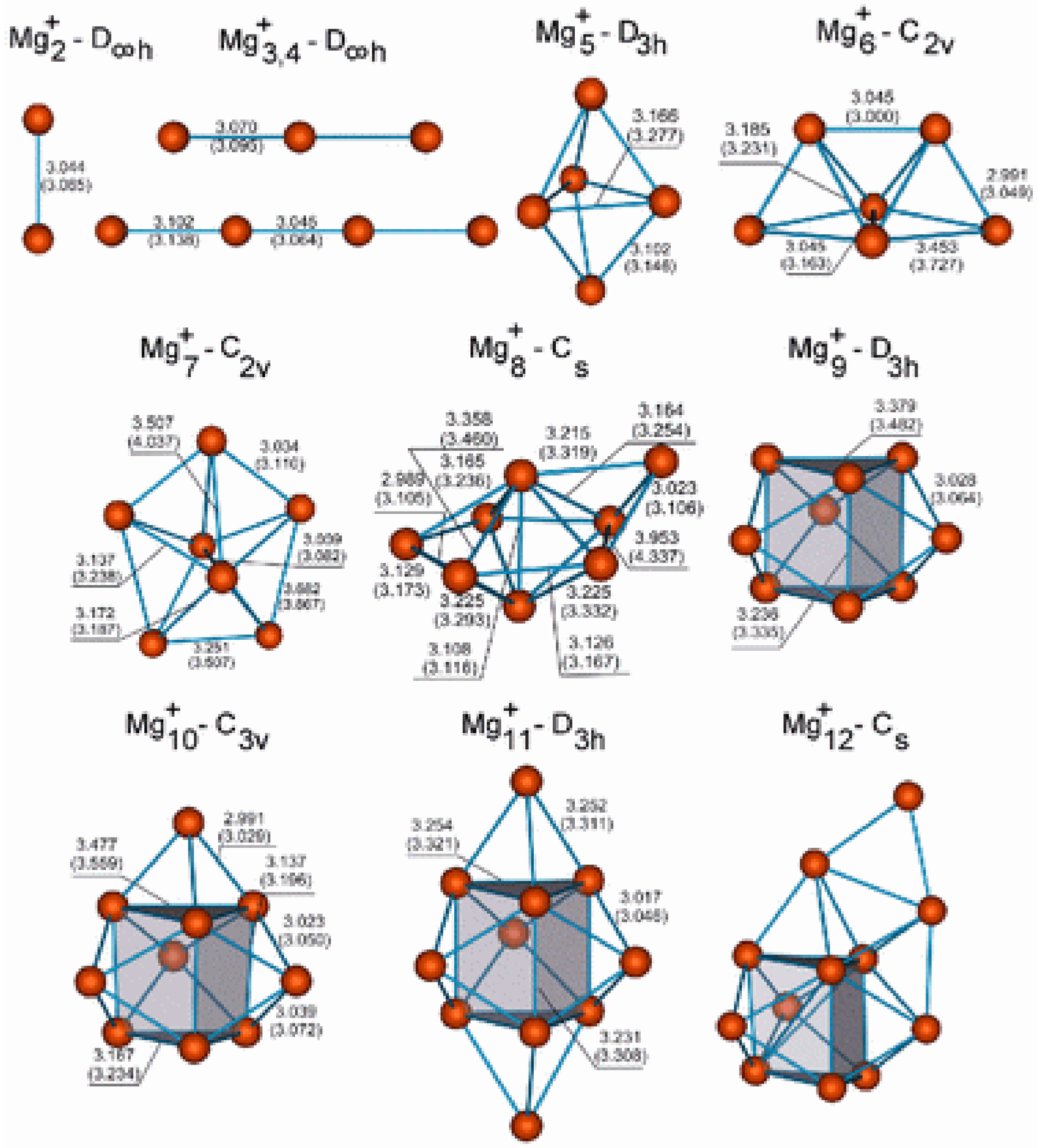}
\includegraphics[scale=0.4]{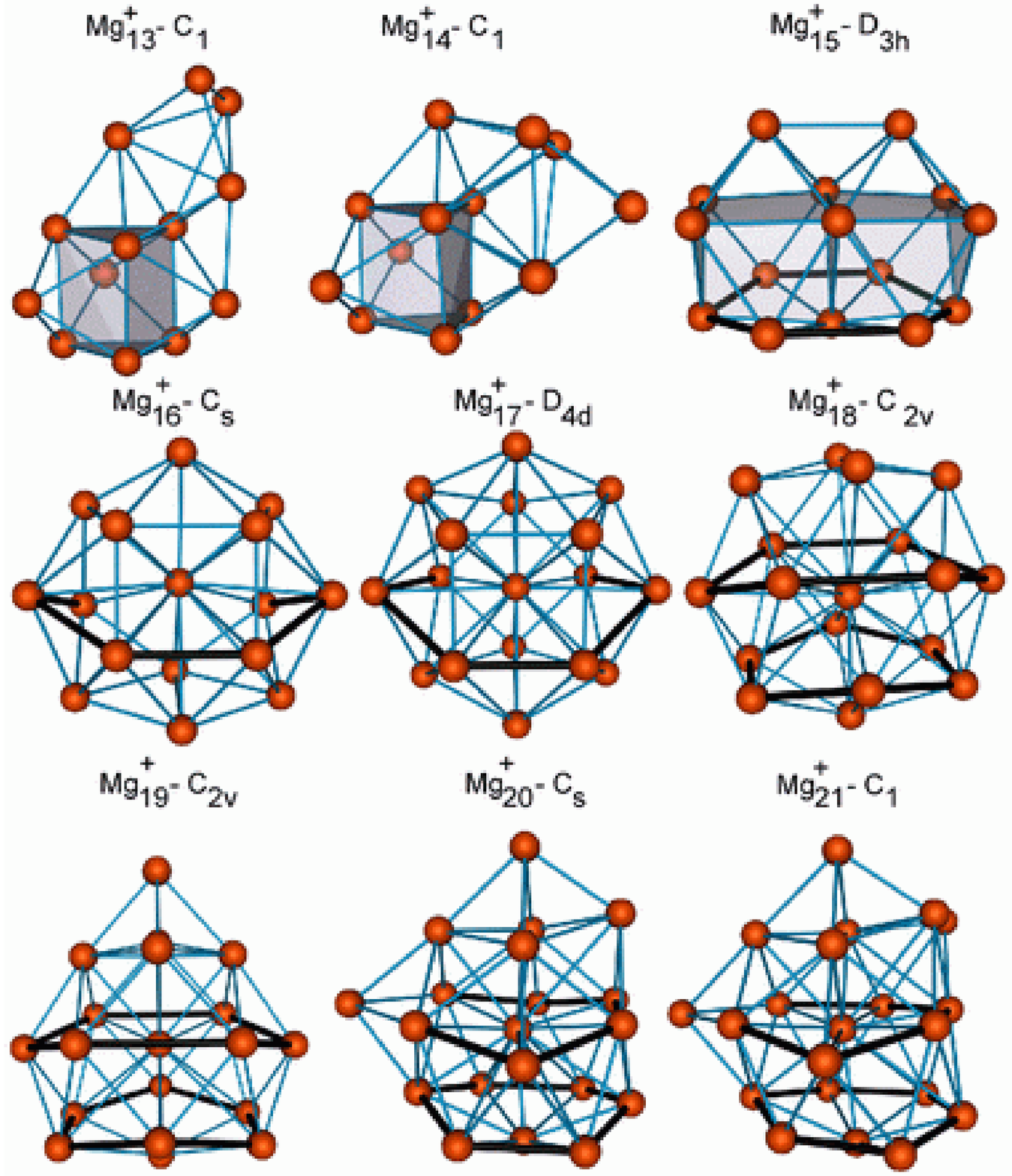}
\caption{
The same as in Fig.\ref{geom_neutral} but for singly-charged magnesium clusters
$Mg_{2}^{+} - Mg_{21}^{+}$.}
\label{geom_ion}
\end{figure}

The optimization of the cluster geometries has been performed with the use of 
the $B3PW91$ and $B3LYP$ methods. For small magnesium clusters with number of atoms 
$N \le 11$, we have also used the {\it ab initio} $MP4$ method in addition to
density-functional calculations. 
With the growth cluster size the {\it ab initio}
$MP4$ calculations become more and more computer time demanding,
therefore we have not performed such calculations for magnesium clusters with 
the number of atoms $N \ge 12$.  
The detail comparison of the results 
obtained by the density-functional and {\it ab initio} perturbation theory methods 
as well as their comparison with the results of other works
is given below, see section \ref{electronic_properties}.
This comparison allows us to conclude that for magnesium clusters the 
$B3PW91$ method is more reliable and 
accurate in comparison with the $B3LYP$ one.   

The results of the cluster geometry optimization for neutral and
singly charged  magnesium clusters consisting of up to 21 atoms are shown in
figures \ref{geom_neutral} and \ref{geom_ion}  respectively.

Magnesium clusters possess various isomer forms those number
grows dramatically with increasing cluster size. In figures 
\ref{geom_neutral} and \ref{geom_ion}, we present only the lowest energy configurations 
optimized by the  $B3PW91$ method. 
The interatomic distances are given in angstroms.  The values in brackets
correspond to the interatomic distances obtained by the $B3LYP$ method.

Figure \ref{geom_neutral} shows that the neutral magnesium clusters form the
compact structures, maximizing the coordination number.
The $Mg_2$ dimer is weekly bound possessing the binding energy per atom
0.039 eV/atom and the bond length 3.609 \AA, which is in a good
agreement with the experimental results of Ref. \cite{Huber_and_Herzberg}, where
the values 0.025 eV/atom for the binding energy and  3.89 \AA\ and for the bond length 
have been reported. 
The lowest energy state for $Mg_3$ is the equilateral triangle, and for 
$Mg_4$ is a regular tetrahedron. 
As we discuss below, the $Mg_4$ cluster is relatively more stable and compact,
as compared to the neighbouring clusters. The $Mg_5$ cluster has a structure of 
slightly elongated triangular bipyramid, while $Mg_6$ consists of three pyramids
connected by their faces, $Mg_7$ is a pentagonal bipyramid, and $Mg_8$ is a 
capped pentagonal bipyramid. These geometrical structures are in a good agreement with the 
results of Ref. \cite{Kumar91}. 

It is worth to note that the optimized geometry structures for small neutral 
magnesium clusters differ significantly from those obtained for sodium clusters
(see, e.g., \cite{Martins85,Bonacic88,StructNa} and references therein). 
Thus, the optimized sodium clusters with $N \leq 6$ have the plane structure. For $Na_6$, 
both plane and spatial isomers with very close total energies exist. 
The planar behavior of small sodium clusters has been explained as a result of the 
successive filling of the  $1\sigma$ and $1\pi$ symmetry orbitals  by
delocalized valence electrons \cite{Martins85},
which is fully consistent with the deformed jellium model calculations \cite{MLSSG02}.
Contrary to the small sodium clusters, the magnesium clusters are tri-dimensional already at
$N=4$, forming the structures nearly the same as the van der Waals bonded clusters.

Starting from $Mg_{9}$ a new element appears in the magnesium cluster structures. 
This is the six atom trigonal prism core, which is marked out in figure \ref{geom_neutral}.
The formation of the trigonal prism plays the important role in the magnesium cluster 
growth process. Adding an atom to one of the triangular faces of the trigonal prism of 
the $Mg_{9}$ cluster results in the $Mg_{10}$ structure, while adding an atom to 
the remaining triangular face of the prism within the $Mg_{10}$ cluster leads to the structure
of $Mg_{11}$, as shown in figure \ref{geom_neutral}.

Further growth of the magnesium clusters for $12 \le N \le 14$ 
leads to the formation of the low symmetry ground state cluster. 
In spite of their low symmetry, 
all these clusters have 
the triagonal prism core. The structural rearrangement occurs for the $Mg_{15}$ cluster, 
which results in the high symmetry structure of the two connected $Mg_{9}$ clusters. 

\begin{figure}[h]
\includegraphics[bb=15 414 628 851, scale=0.6]{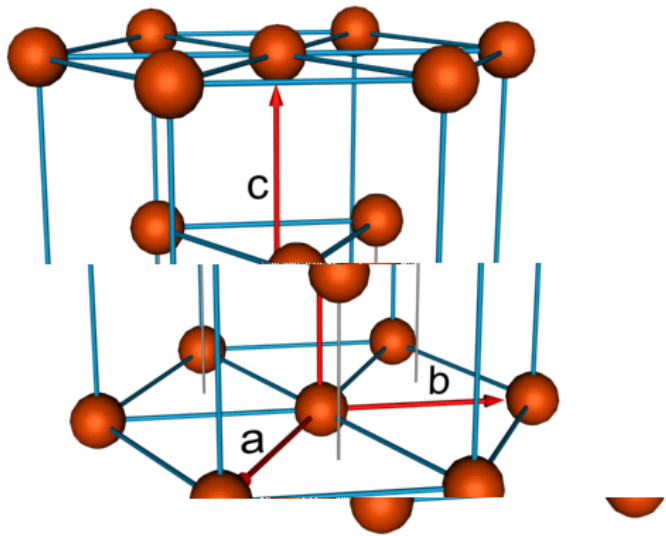}
\caption{Primitive cell for magnesium {\it hcp} lattice. 
For bulk magnesium $a=b=3.21$ \AA\ and $c=5.21$ \AA\ \cite{Ashcroft}.}
\label{primitive_cell}
\end{figure}

Starting from $Mg_{15}$ another motif based on the hexagonal ring structure 
which is marked out in  figure \ref{geom_neutral}
dominates the cluster growth. Such a ring is the basic element of the 
hexagonal closest-packing ({\it hcp}) lattice,
as one can see in figure \ref{primitive_cell}, in which the primitive cell for the
magnesium {\it hcp} lattice is presented. 
Thus, $N=15$ is the turning-point 
in the formation of the {\it hcp} lattice for magnesium.

Vectors ${\bf a}$, ${\bf b}$ and ${\bf c}$ in figure \ref{primitive_cell} show 
the primitive cell axes of the {\it hcp} lattice.  
For bulk magnesium $a=b=3.21$ \AA\ and $c=5.21$ \AA\ \cite{Ashcroft}.
The fundamental characteristic for
the hexagonal closest-packing of spheres is the value of ratio $c/a$, 
which is equal to $\sqrt{8/3} \approx 1.633$ for ideal {\it hcp}  lattice. 
The bulk magnesium with $c/a = 1.62$ is very close to ideal 
{\it hcp} structure \cite{Kittel}.

The distinct three-layered structure of $Mg$-clusters with $N \ge 18$ based on the 
hexagonal rings allows one to determine
the averaged values of the primitive axes $\langle c \rangle$ and $\langle a \rangle$. 
Table \ref{tab:Mg_cell} demonstrates that the calculated values 
$\langle c \rangle$ and $\langle a \rangle$
and their ratio for magnesium clusters with 
$N \ge 18$ are very close to the corresponding values for bulk magnesium.

\begingroup
\begin{table*}[h]
\caption{The averaged values of the primitive axes and its ratio for the 
{\it hcp} lattice element for magnesium clusters with $N \ge 18$ calculated within the $B3PW91$
approximation. Values in brackets correspond to singly-charged magnesium clusters.}
\label{tab:Mg_cell}

\begin{ruledtabular}
\begin{tabular}{ccccccc}
 
     \multicolumn{1}{c}{  } &
     \multicolumn{1}{c}{$Mg_{18}$} &
     \multicolumn{1}{c}{$Mg_{19}$} &
     \multicolumn{1}{c}{$Mg_{20}$} &
     \multicolumn{1}{c}{$Mg_{21}$}&
     \multicolumn{1}{c}{Mg bulk, \cite{Ashcroft} }\\ 

\hline 

$\langle c \rangle$, \AA                & 5.08 (5.42)& 5.47 (5.37)& 5.48 (5.23)& 5.56 (5.23) & 5.21 \\
$\langle a \rangle$, \AA                & 3.14 (3.19)& 3.05 (3.22)& 3.20 (3.21)& 3.20 (3.21) & 3.21 \\
$\langle c \rangle / \langle a \rangle$ & 1.62 (1.70)& 1.79 (1.67)& 1.71 (1.63)& 1.74 (1.63) & 1.62 \\

\end{tabular}
\end{ruledtabular}
\end{table*}
\endgroup

Figure \ref{geom_ion} shows the 
optimized geometries of singly-charged cationic magnesium clusters. 
The ground state geometries of the cationic magnesium clusters are not very different
from those obtained for the neutral parent clusters with the exception 
of $Mg_{3}^{+}$ and $Mg_{4}^{+}$, the  equilibrium geometries of which are linear chains.
Below, we discuss the stability of the linear chain isomers 
for the magnesium clusters (neutral and singly-charged) within the size range considered.   

\begin{figure}[h]
\includegraphics[scale=1.3]{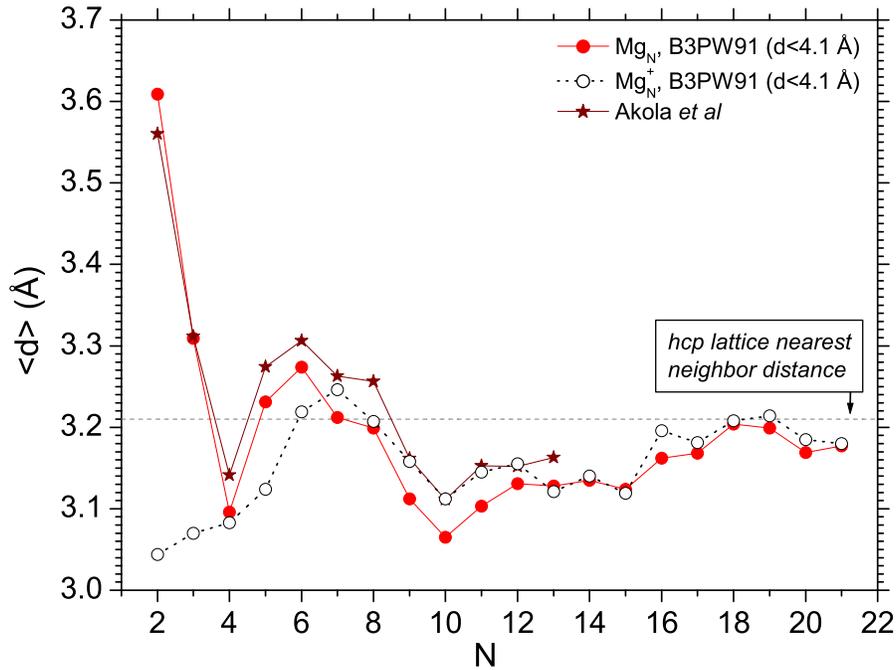}
\caption{The average bonding distance as a function
of cluster size for neutral and singly-charged magnesium clusters.
Stars present the results of the work by Akola {\it et al} \cite{Akola01}.
The horizontal dashed line indicates the bulk limit for the {\it hcp} lattice \cite{Ashcroft}.}
\label{dist_Rav}
\end{figure}

In figure \ref{dist_Rav}, we present the 
average bonding distance, $\langle d \rangle$, calculated within the $B3PW91$ approximation for 
neutral and singly-charged magnesium clusters.
When calculating the average bonding distance in a cluster,
interatomic distances smaller than  4.1 \AA\,\, have only been taken into account.
The bulk limit for the magnesium {\it hcp} lattice \cite{Ashcroft} indicated in figure 
by horizontal dashed line.

Figure \ref{dist_Rav} shows how the average bonding distance 
evolve with increasing cluster size. 
It is clearly seen that the dependence of the average bonding distance
on cluster size has essentially non-monotonous oscillatory behavior. 
For $Mg_2$, the bonding distance calculated within the $B3PW91$ method is equal 
to 3.609 \AA,\, which is in a good agreement with the experimental result 
3.891 \AA\, of Ref. \cite{Huber_and_Herzberg}. The appearance of the minima 
in the size dependence of the average bonding distance 
shows that $Mg_{4}$, $Mg_{10}$, and $Mg_{20}$ clusters (8, 20 and 40 valence electrons, 
respectively) are more tightly packed than their neighbours. 
This behavior can be interpreted by the influence of electronic shell effects 
on the geometrical structure of magnesium clusters. It supports the conclusion 
of Ref. \cite{Reimann97} that electronic shell effects can 
enhance the stability of geometric structures resulting from dense ionic packing.

Additional minimum in the dependence of the average bonding distance on $N$
arrises at $N = 15$. At this $N$ a considerable
rearrangement of the cluster geometry take place as it is seen from 
figure \ref{geom_neutral}. 
Indeed, starting from the $Mg_{15}$ 
cluster the three-layered 
structure based on the hexagonal ring is formed. 
It is clearly seen in figure \ref{dist_Rav} 
that for $N \ge 15$ the average bonding distance for magnesium clusters 
approaches the bulk limit.

The evolution of the average bonding distance with cluster size differs for
magnesium clusters from that for sodium.
For neutral sodium clusters, one can see
odd-even oscillations of $\langle d \rangle$ atop its systematic
growth and approaching the bulk limit \cite{StructNa}. These features
have the quantum origin and arise due to the spin coupling of the 
delocalized valence electrons.
For magnesium clusters, the average bonding distance depends on size non-monotonically, with 
minima for the $Mg_{4}$, $Mg_{10}$, $Mg_{15}$, and $Mg_{20}$ clusters.
Such an irregular behavior is induced by both the closure of electronic
shells of the delocalized electrons and structural rearrangements.

Manifestation of the magic numbers  in the 
dependence of the average bonding distance on cluster size
coinciding with the spherical jellium model magic numbers does not imply, however,  
the rapid metallization of magnesium clusters. To investigate the transition of 
van der Waals to metal bonding in magnesium clusters it is necessary to explore
in detail the evolution of their electronic properties. Below we perform such analysis
in detail. 

Dashed line in figure \ref{dist_Rav} shows the average bonding distance 
as a function of cluster size calculated for singly-charged magnesium clusters. 
Figure \ref{dist_Rav} demonstrates the essential difference in the cluster size dependence of
$\langle d \rangle$ for the cationic and neutral magnesium 
clusters with $N \le 6$.
The small cationic magnesium 
clusters are more compact in comparison with the corresponding neutral clusters. 
For example, for $Mg_{2}^{+}$ the bonding distance is equal to 3.044 \AA,\, 
which is much less than in the case of $Mg_{2}$. This phenomenon has a simple 
physical explanation: the removed electron is taken from the antibonding orbital. 
The fact that cationic magnesium clusters are more stable
than the parent neutral and anionic clusters has been already noted in \cite{Reuse90}.

Within the size range $N \ge 7$,  the average bonding distances
for single-charged  and neutral magnesium clusters behave similarly.
The absolute value of $\langle d \rangle$ for single-charged clusters 
is slightly larger in this region of $N$.

Figure \ref{dist_Rav} demonstrates the good agreement of our results with the dependence
of  $\langle d \rangle$ on $N$ calculated in \cite{Akola01} for neutral $Mg$-clusters
within the size range $N \le 13$.

\subsection{Binding energy per atom for Mg$_{N}$ and Mg$_{N}^{+}$ clusters.}
\label{electronic_properties}

The binding energy per atom for small neutral and
singly-charged magnesium clusters is defined as follows:
\begin{eqnarray}
E_b/N&=&E_1-E_N/N 
\label{E_b}
\\
E_b^+/N&=& \left((N-1)E_1+E_1^+-E_N^+\right)/N,
\label{E_b^p}
\end{eqnarray}
\noindent
where $E_N$ and $E_N^+$ are the energies of a neutral  and
singly-charged N-particle atomic cluster,
respectively. $E_1$ and $E_1^+$ are the energies of a single magnesium atom and an ion.

\begin{figure}[h]
\includegraphics[scale=1.3]{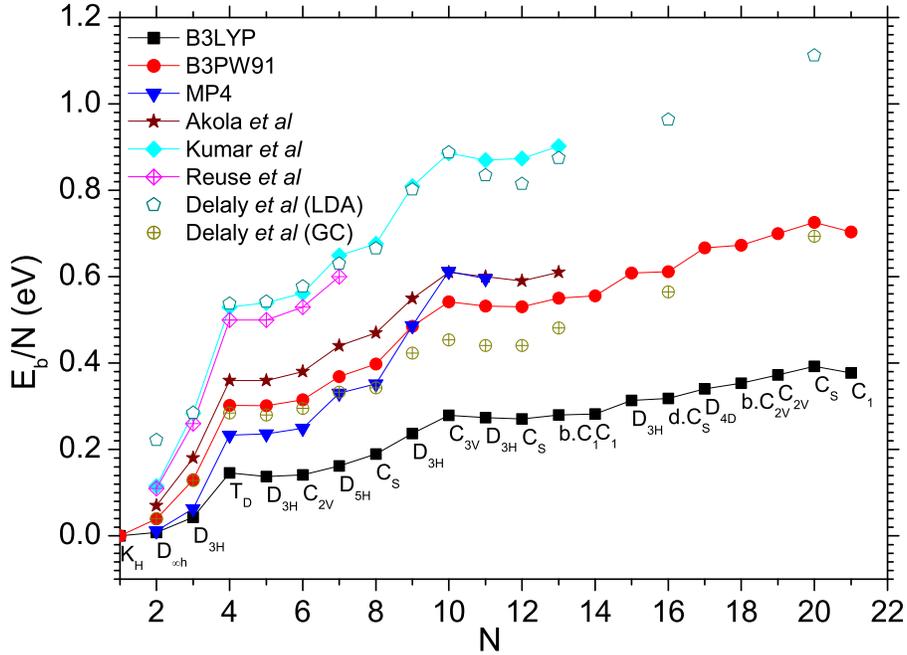}
\caption{Binding energy per atom for neutral magnesium clusters as a function of
cluster size.
Squares, circles and lower triangles
 represent the binding energies per atom calculated
by the $B3LYP$,  $B3PW91$ and $MP4$ methods respectively. 
Stars, filled rhombus  and crossed rhombus
show the results of the works 
by Akola {\it et al} \cite{Akola01}, Kumar {\it et al} \cite{Kumar91}, and  
 Reuse {\it et al} \cite{Reuse90} respectively.
Opened pentagons and crossed circles show the result of Delaly {\it et al} \cite{Delaly92}
obtained with the use of the 
LDA and gradient-corrected (GC) LDA methods respectively.
Labels indicate the point symmetry group of the isomers represented.
Their geometries one can find in section \ref{geom_opt}.}
\label{binding_neutral}
\end{figure}

\begin{figure}[h]
\includegraphics[scale=1.3]{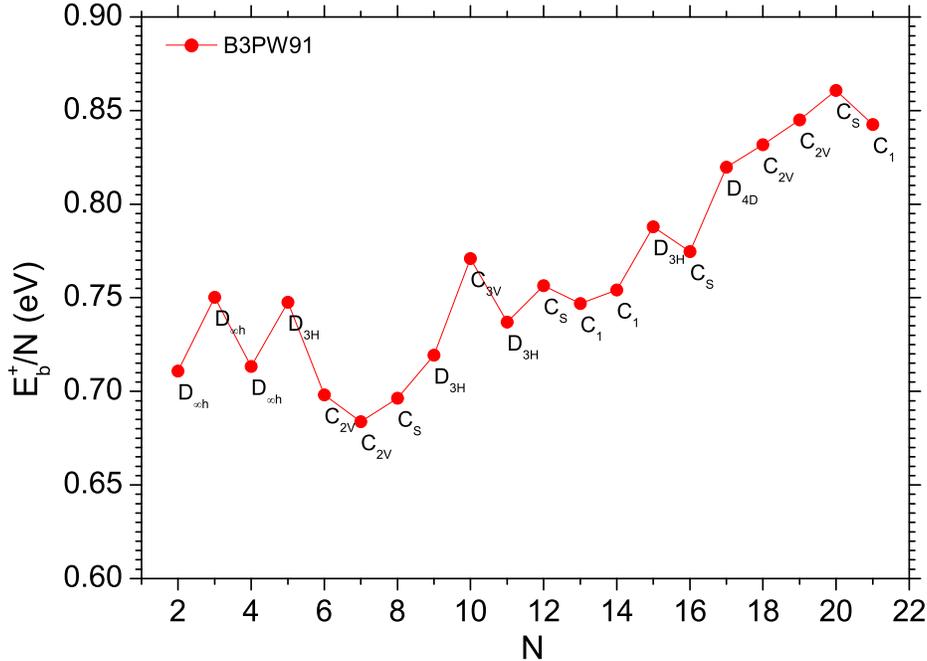}
\caption{The same as in Fig.\ref{binding_neutral} but for 
singly-charged magnesium clusters.}
\label{binding_ion}
\end{figure}

Figures \ref{binding_neutral} and \ref{binding_ion} show the
dependence of the binding energy per atom for neutral and
singly-charged clusters as a function of cluster size.
The energies of clusters have been obtained
using the $B3LYP$, $B3PW91$ and $MP4$
methods.
Calculations of the binding energies have
been performed by different theoretical methods and with the use of different 
exchange-correlation functionals for the sake
of comparison of their accuracy and computation efficiency.
In figure \ref{binding_neutral} filled rhombus, crossed rhombus and  opened pentagons
show the result of calculations by Kumar {\it et al} \cite{Kumar91},
Reuse {\it et al} \cite{Reuse90} and Delaly {\it et al} \cite{Delaly92} respectively. 
These calculations have been performed within
the Hohenberg-Kohn-Sham local-density approximation using the Perdew and Zunger 
\cite{Perdew_Zunger_81} parameterization of the Ceperley and Alder \cite{Ceperley_Alder_80}
data for the exchange correlations.  
Crossed circles and stars present the results of 
Delaly {\it et al} \cite{Delaly92} and Akola {\it et al} \cite{Akola01} derived with the use of
the gradient-corrected approximation \cite{Perdew86,Car_Parrinello_85} and
the PBE parameterization 
of the gradient-corrected exchange-correlation energy functional \cite{Perdew96} 
respectively. 
 
Figure \ref{binding_neutral} shows that, although, the qualitative behavior of the binding 
energy per atom calculated within different approaches is similar, the quantitative discrepancy
between the curves is rather considerable. 
This is a result of different accounting for 
the gradient corrections  to the local-density
exchange correlation interaction within different  methods.
The gradient corrections have been shown to provide a systematic improvement
in the computed properties of magnesium clusters \cite{Delaly92}. 
The difference in the binding 
energy per atom for neutral magnesium clusters with  $N \le 21$ 
calculated with the use of the 
gradient corrected $B3LYP$ and $B3PW91$ methods reaches 0.35 eV. 
The reason for this difference is in the different way of accounting for many-electron 
correlations within the $B3LYP$ and $B3PW91$ methods. To explore what type of 
parameterization of the exchange-correlation energy is more reliable for magnesium
clusters we have used the post-Hartree-Fock 
{M\o ller}-Plesset perturbation theory.
This method is free of phenomenological parameters and can be
used as a criterion for checking the reliability of various density-functional theory schemes. 
The disadvantage of the perturbation theory approach consists in the fact that it leads to
the dramatic growth of the 
computational costs with increasing the number of electrons in the system 
in comparison with that for the density-functional theory calculations.
Therefore, we have used the $MP4$ method only for clusters with the number of atoms 
$N \leq 11$.

Figure \ref{binding_neutral} shows that the results of the $MP4$ theory 
are in a reasonable agreement with those derived by the $B3PW91$ method. This comparison 
demonstrates that for magnesium clusters simulations the
$B3PW91$ method is more reliable than 
the $B3LYP$ one. Our results derived within the $B3PW91$ and $MP4$ 
approximations are
in a good agreement with those from Ref. \cite{Delaly92,Akola01}.

\begin{figure}[h]
\includegraphics[scale=1.3]{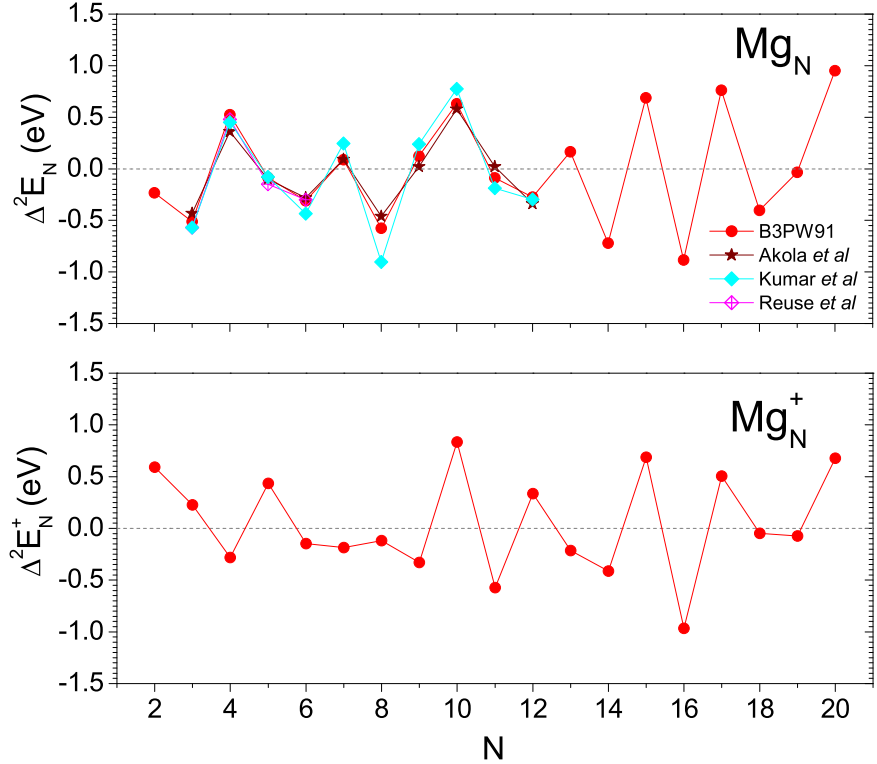}
\caption{Second differences of total energy for 
neutral, $\Delta^2E_{N}^{}=E_{N+1}^{} - 2E_{N}^{} + E_{N-1}^{}$, 
and singly-charged, $\Delta^2E_{N}^{+} = E_{N+1}^{+} - 2E_{N}^{+} + E_{N-1}^{+}$, 
magnesium clusters. 
Stars show the result of the work by Akola {\it et al} \cite{Akola01},
filled rhombus by Kumar {\it et al} \cite{Kumar91}, and  
crossed rhombus by Reuse {\it et al} \cite{Reuse90}.}
\label{der_binding}
\end{figure}

We now discuss the behavior of the binding energy as a function of cluster size 
for both neutral and singly-charged magnesium clusters.
For neutral magnesium clusters, the binding energy 
per atom increases steadily with the growth cluster size. The local maxima of $E_b/N$ at
$N=$ 4, 10 and 20 correspond to the most stable configurations of the magic
magnesium clusters possessing 
$N_{el}=$ 8, 20 and 40 valence electrons respectively. 
This behavior is in agreement with the simple spherical jellium model. 
The analysis of the second differences of the binding energy 
(see Fig.\ref{der_binding}) 
confirms this conclusion and makes a hint about relative stability 
of the $Mg_{7}$, $Mg_{13}$, $Mg_{15}$, and $Mg_{17}$ 
clusters, in addition to the magic clusters $Mg_{4}$, $Mg_{10}$ and $Mg_{20}$ .
The additional magic numbers can be explained within the deformed jellium model 
accounting for spheroidal deformations of the cluster core 
(see, e.g., \cite{LSSCG00,Lyalin01a,MLSSG02} and references therein). 
For a spheroidal jellium cluster, 
the orbital angular momentum does not remain
a good quantum number characterizing the valence electrons energy levels.
In this case, the energy levels are characterized by
the projection of the angular momentum $\Lambda$
on the principal axis and by the parity of the wave function.
Thus, the energy levels with $\Lambda=0$ are twofold degenerated on the projection of
the electron's spin, while those
with $\Lambda\neq 0$ are fourfold degenerated both on the projection 
of the electron spin and on the sign of the projection $\Lambda$ on the principal 
cluster axis.
The deformed jellium clusters having closed electronic subshells
possess the enhanced stability. 
Therefore, in addition to the spherical magic clusters with 8, 20, 40 etc valence electrons,
the deformed jellium clusters with 6, 10, 14, 18, 22, 26, 30, 34 etc. valence electrons 
turn out to be relatively stable. This fact leads to the following additional magic numbers
3, 5, 7, 9, 11, 13, 15, 17 for the jellium magnesium clusters. Some of these 
numbers, such as 3, 5, 9, 11 precede or follow the spherical magic numbers 4, 10, 20 
and as a result of that become masked and are not that pronounced in the second differences 
analysis. 

For singly-charged magnesium clusters, the binding energy per atom as a function of cluster 
size is essentially non-monotonous.
The local maxima of the binding energy
for the $Mg_{3}^{+}$, $Mg_{5}^{+}$, $Mg_{10}^{+}$, $Mg_{12}^{+}$, $Mg_{15}^{+}$ 
and $Mg_{20}^{+}$ clusters indicate their enhanced stability.  
Figure \ref{der_binding} shows second differences of the total 
energy for singly-charged magnesium clusters.
This figure demonstrates the enhanced stability of the mentioned cluster ions 
and the $Mg_{17}^{+}$ cluster.

The sequence of magic numbers for singly charged magnesium clusters differs 
from that for neutral clusters. This happens because singly charged magnesium clusters
always possess odd number of valence electrons and, thus, always contain open 
electronic shells. For neutral magnesium clusters, situations of both close and open 
electronic shells are possible. The enhanced stability of a $Mg$-cluster ion arises, when 
its electronic configuration has one hole in or an extra electron above the filled shells.
Thus, the cluster ions $Mg_{5}^{+}$, $Mg_{11}^{+}$ and $Mg_{21}^{+}$ contain one extra
electron over the complited spherical electronic shells, while the clusters
$Mg_{4}^{+}$, $Mg_{10}^{+}$ and $Mg_{20}^{+}$ have a hole in the spherical 
outer electronic shell. Our results presented in figures \ref{binding_ion} 
and \ref{der_binding} demonstrate that the cluster ions $Mg_{5}^{+}$,  $Mg_{10}^{+}$ and
$Mg_{20}^{+}$ turn out to be more stable than their neighbors. We note that the 
alteration of the magic number from $N = 4$ for neutral $Mg$-clusters to 
$N = 5$ for $Mg$-cluster ions happens
because the electronic configuration containing an extra electron becomes more favorable for   
$Mg_{5}^{+}$. This is not the case for the $Mg_{10}^{+}$ and $Mg_{20}^{+}$ clusters,
those outer electronic configurations contain a hole.

The $Mg$-cluster mass spectra have been recorded in \cite{Diederich01} indicating 
the enhanced stability of the clusters with $N = $ 5, 10, 15, 18 and 20. In that 
work the role of the cluster ionization was not reliably clarified \cite{Diederich01}
and thus the charge state of the clusters was not reliably determined.
As a result, the observed magic numbers sequence should be a combination of the
magic numbers sequences for neutral and singly-charged cluster ions.
Thus, $N=$ 5 is the ionic magic number, $N=$ 10, 15 and 20 are the magic numbers
manifesting themselves clearly for both neutral $Mg$-clusters and $Mg$-cluster ions.
The second differences are positive and relatively large for $N=$ 13 (neutral clusters)
and $N=$ 12 (singly-charged cluster ions). Possibly, the interplay between neutral
clusters and ions make these numbers masked in experiment.
The second differences are also positive for $N=$ 7 for neutral $Mg$-clusters
and for $N=$ 3 for $Mg$-cluster ions, although the enhancement for those
numbers have not been experimentally observed. We explain this fact by possible
suppression of the experimental signal in the region of small $N$ and relatively
small values of the second differences in the mentioned cases.

\begin{figure}[h]
\includegraphics[scale=0.7]{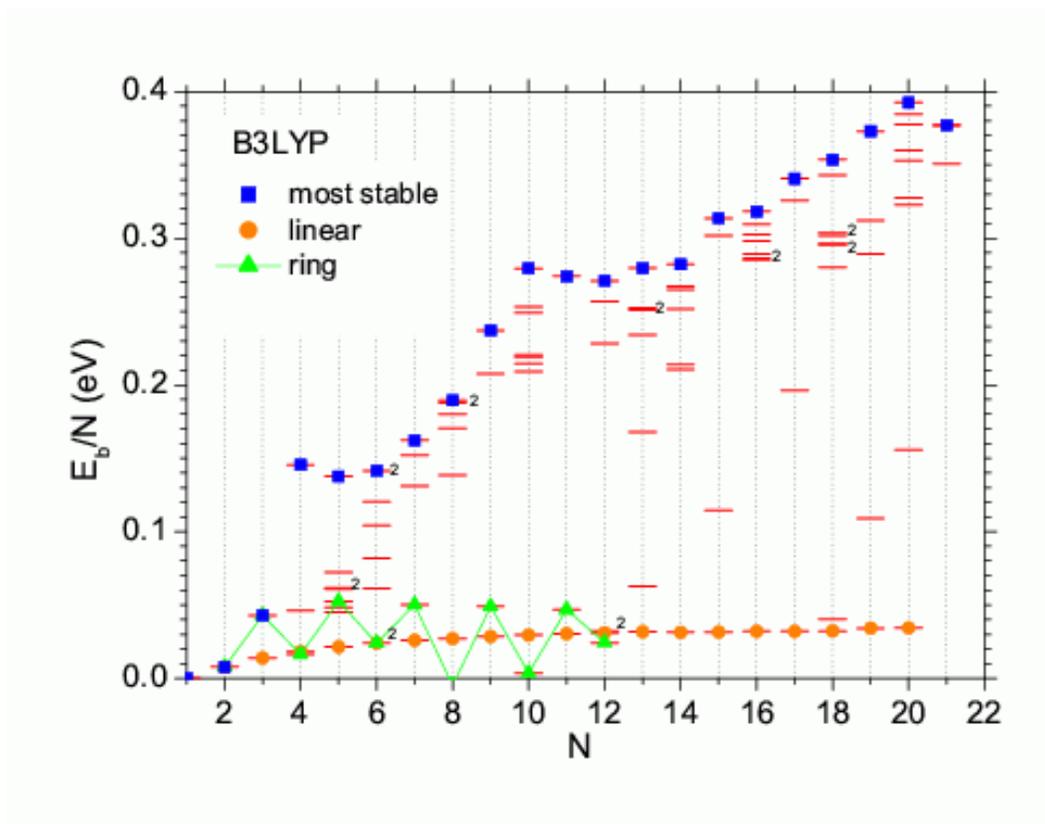}
\vspace*{-3cm}
\includegraphics[scale=0.7]{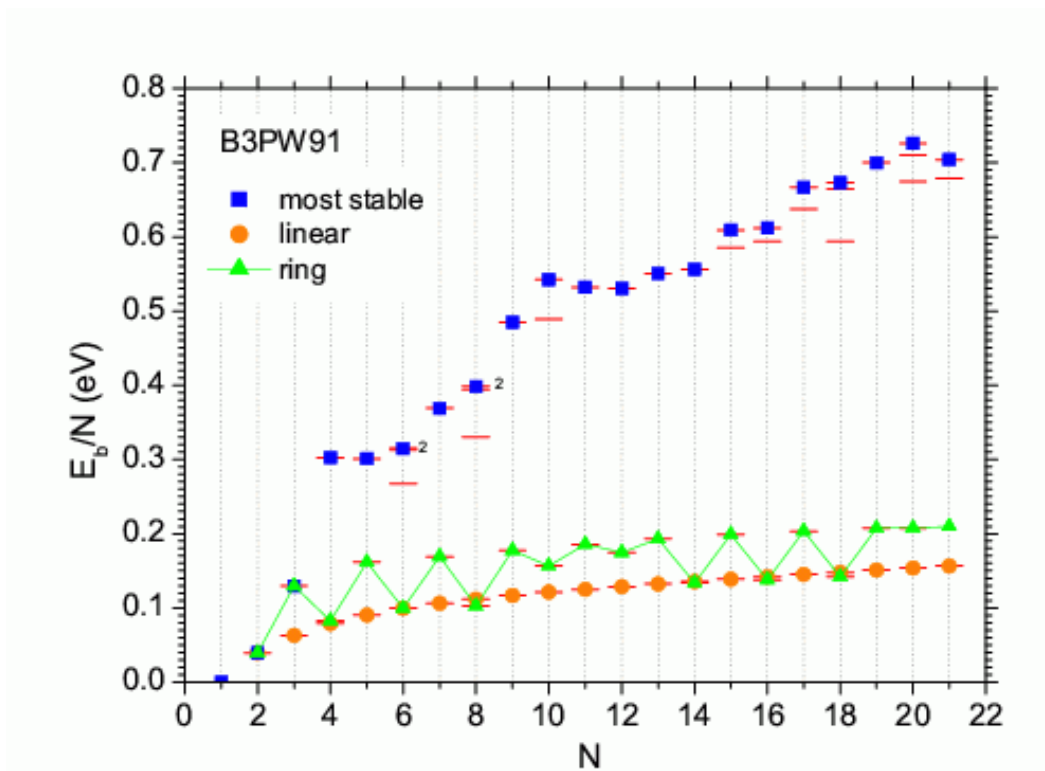}
\caption{Binding energy per atom for a variety of isomers of  
neutral magnesium clusters as a function of cluster size.
The corresponding point symmetry groups and the 
accurate values of the total energies
are presented in Appendix in tables \ref{tab:B3LYP} and \ref{tab:B3PW91}.
Numbers near some lines show the number of found isomers with the corresponding
close energies.}
\label{Eb_spectra}
\end{figure}

The potential energy surface for a cluster becomes more and more complicated
with increasing cluster size. The magnesium clusters are not an exception.
Figure \ref{Eb_spectra} demonstrates this fact where we present
the binding energies per atom calculated for a variety of isomers of neutral magnesium clusters.    
The corresponding point symmetry groups and the accurate values of the total energies
calculated within the $B3LYP$ and $B3PW91$ approximations
are presented in Appendix in tables 
\ref{tab:B3LYP} and \ref{tab:B3PW91} respectively.
Most of the isomer configurations have been obtained using the $B3LYP$ method,
while the $B3PW91$ method has been used for the exploration of the ground state 
energy isomers, as well
as for the linear and ring-like isomer structures.

Squares in figure \ref{Eb_spectra} correspond to the most stable clusters possessing the 
minimal total energy. Among the variety of isomers, presented in figure
\ref{Eb_spectra}, we mark certain groups of isomers with the 
fixed symmetry. So, circles present the linear chains ($D_{\infty h}$ 
point symmetry group) and the upper triangles correspond to the rings of 
$N$ atoms ($D_{N h}$ point symmetry group). It is an interesting fact that among the 
multitude of the isomers of neutral magnesium clusters the linear chains and rings 
are always stable.   
We pay a particular attention to these structures
because of their possible applications in nano-technology. 
Extracting these isomers and putting them on a substrate one can produce one-atom 
wide quantum wires.
The linear chains and rings of atoms
are also very interesting 
from the theoretical point of view, because with their help one can investigate 
the transition from 
one- to two-dimensional structures.
For linear chains, the binding energy per atom increases slowly with the growth the 
number of atoms,
while in the case of rings the value of $E_{b}/N$ has the prominent 
odd-even oscillatory behavior.

This behavior arises as a result of successive filling the $\sigma$- and $\pi$-symmetry orbitals 
by valence electrons in  magnesium linear chains and rings. 
Indeed, because of its symmetry the one-dimensional linear chain 
of $N$ magnesium atoms has the following configuration of valence electrons: 
$1\sigma^{2}$, $2\sigma^{2}$,  $3\sigma^{2}$, ..., $N\sigma^{2}$. 
Therefore for any $N$ it has the closed electronic shell structure. 
This fact explains the monotonous growth with $N$ of the linear chain binging energy, 
and its relative saturation in the region $N > 10$.  

The molecular orbitals for the structure of the ring-type have to be aligned with the plane 
of the ring. Such orbitals are fourfold degenerated due to symmetry reasons.
The $Mg_{2}$ dimer has four valence electrons that occupy  
spherically-shaped $1\sigma^{2}$ and prolate-like $2\sigma^{2}$ orbitals.
The $Mg_{3}$ trimer has six valence electrons, two of them occupy $1\sigma^{2}$ state,
while the remaining four electrons fill the fourfold degenerated  $1\pi^{4}$ orbital aligned
with the plane of the trimer. With increasing the number of magnesium atoms in the ring, 
the valence electrons continue to occupy fourfold degenerated orbitals aligned with 
the plane of the ring. Therefore, in the magnesium ring-like isomers with the odd number of atoms 
all electronic shells are closed, while the isomers with the even number of atoms have the open
electronic shell. This fact results in the enhanced stability of the  magnesium rings 
with an odd number of atoms $N=3, 5, 7, 9, ...$ etc. and explains the odd-even 
oscillatory behavior of the binding energy for the magnesium rings.

\subsection{Ionization potentials and HOMO-LUMO gaps}

\begin{figure}[h]
\includegraphics[scale=1.3]{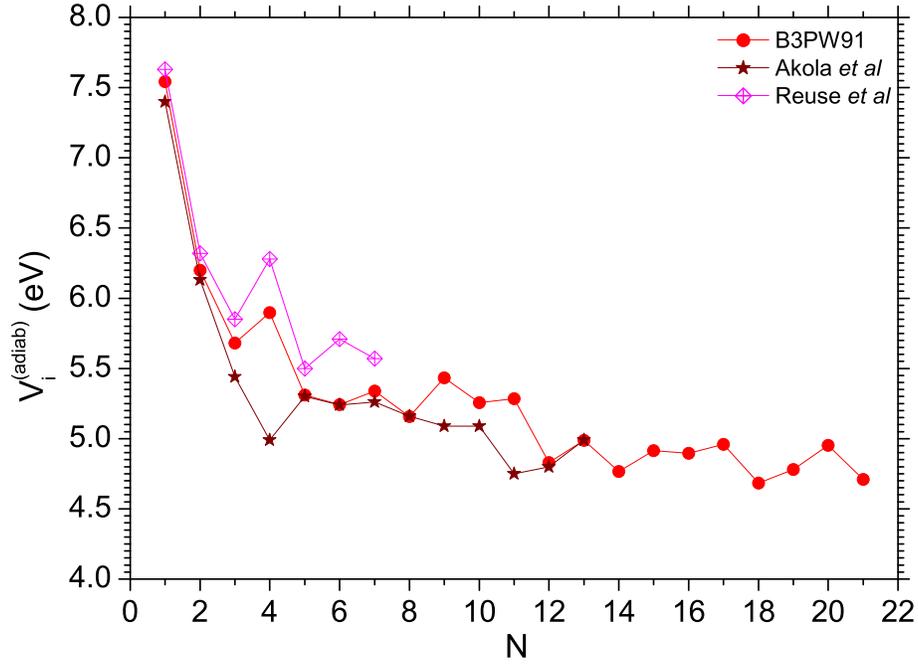}
\includegraphics[scale=1.3]{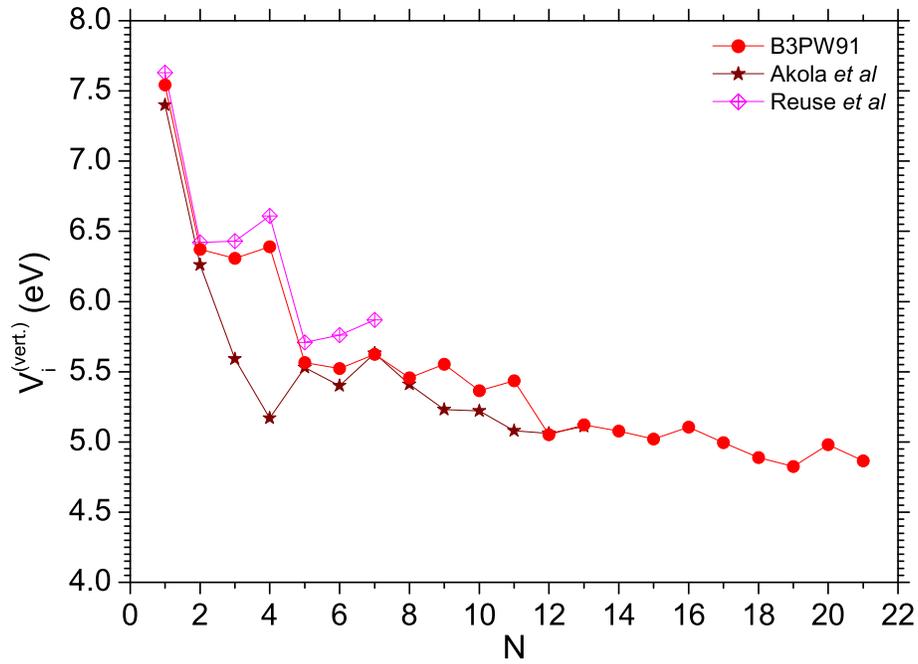}
\caption{Adiabatic, $V_{i}^{adiab}$, and vertical, $V_{i}^{vert}$, 
ionization potential for small magnezium clusters.
Stars and crossed rhombus show the result of the work
by Akola {\it et al} \cite{Akola01} and  by Reuse {\it et al} \cite{Reuse90} 
respectively.}
\label{ionization_potential}
\end{figure}

Let us now consider how the ionization potential of magnesium clusters
evolves with increasing cluster size. 
The ionization potential of a cluster $V_{i}$ is equal to the difference 
between the energies of the corresponding cluster ion and the neutral cluster,
$V_{i} = E_N^+ - E_N$.
Figure \ref{ionization_potential} shows the dependence of the
adiabatic $V_{i}^{adiab}$ (i.e. the geometry relaxation of the ionized cluster 
is taken into account)  
and vertical $V_{i}^{vert}$ (i.e. the cluster geometry is frozen during the ionization process)
ionization potential on $N$.
We compare our results derived by the $B3PW91$ method with theoretical data 
from Ref. \cite{Akola01} and \cite{Reuse90} and with the bulk limit, $V_{i}^{bulk}=3.64$ eV,
taken from \cite{Ashcroft}.

Both the vertical and adiabatic ionization potentials 
evolve non-monotonously 
with increasing cluster size. Figure \ref{ionization_potential} shows 
that ionization potential
of magnesium clusters steadily but rather slow decreases towards the bulk limit.
This evolution is neither rapid nor monotonous process.
In order to exclude the influence of the cluster geometry rearrangement, we first  
consider the vertical ionization potential. The size dependence of the vertical ionization
potential has a prominent maximum at $N=4$ 
followed by a sharp decrease. Such a behavior of the ionization
potential is typical for the jellium model, predicting maxima in the size dependence
of the ionization potential at the magic numbers corresponding to 
the clusters with closed electronic shells.
Our data are in a good agreement with 
the results of Ref. \cite{Reuse90}, but contradict to those reported in 
Ref. \cite{Akola01} for the $Mg_{3}$ and $Mg_{4}$ clusters. 
In \cite{Akola01} the appearance of the deep minimum in vertical ionization potential at $N = 4$  
was explained as a result of a stronger charge delocalization in the $Mg_{4}$ 
cluster in comparison with its neibours.

We note that the peculiarities in the ionization potential dependence on $N$ 
correlate with the magic numbers that appear for the
singly-charged magnesium clusters. Indeed, the minima in $V_{i}^{vert}$
correspond
to the maxima in  $E_{b}^{+}/N$ for $Mg$-cluster ions (see Fig. \ref{binding_ion}).
This fact has a simple explanation. The ionization potential of a cluster is 
equal to the difference 
between the energies of the corresponding cluster ion and the neutral cluster. 
For neutral $Mg$-clusters, the binding energy as a function of $N$ increases steadily with the 
growth cluster size, while for $Mg$-cluster ions - irregularly. 
Thus, their difference mimics all the irregularities that appear in the binding energy
dependence on $N$ for singly-charged magnesium clusters.

For $N \ge 6$, the vertical ionization
potential changes slowly with increasing cluster size. 
This process is characterized by the irregularities
that originate due to the influence of the cluster geometry on the 
jellium-type electronic structure of $Mg$-clusters. 

Indeed, the shape of a jellium cluster is defined by its 
electronic structure. Thus, the closed shell jellium clusters are spherical,
while clusters with opened electronic shells are deformed due to the Jahn-Teller distortions.
The jellium picture works fairly well for sodium clusters. 
The ionization
potential of sodium clusters drops rapidly and systematically at the electronic 
shell closures.
The $N$-dependence of the ionization potential  has prominent, regular
odd-even oscillations (see, e.g., \cite{StructNa,MLSSG02} and references therein).
Magnesium clusters are different.
As discussed in section \ref{geom_opt},
the evolution of the $Mg$-cluster geometry 
is closely connected with the formation of elements of the {\it hcp} lattice cell. 
Although, the electronic shell effects clearly manifest themselves in 
the formation of the $Mg$-cluster geometry, they do not determine it completely.
Another words, there is an interplay of the jellium and 
the {\it hcp} lattice factors in the formation of the $Mg$-cluster geometry and  
the electronic properties such as the binding energy and the  ionization potential.

The adiabatic ionization potential dependence that is shown in figure \ref{ionization_potential}
exhibits qualitatively the same behavior as the vertical one, however, has more 
pronounced irregularities due to geometry rearrangements of the ionized clusters.

\begin{figure}[h]
\includegraphics[scale=1.3]{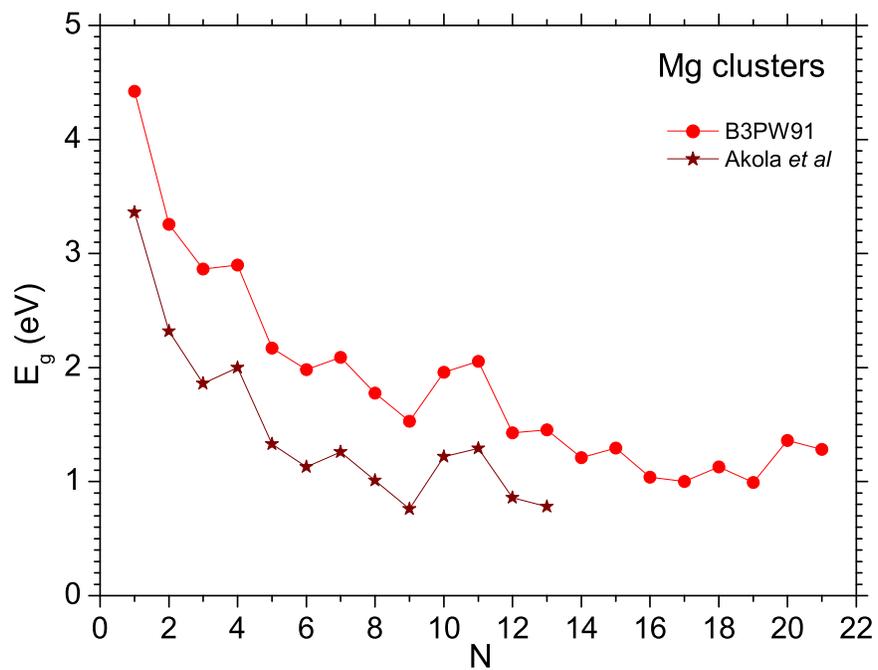}
\includegraphics[scale=1.3]{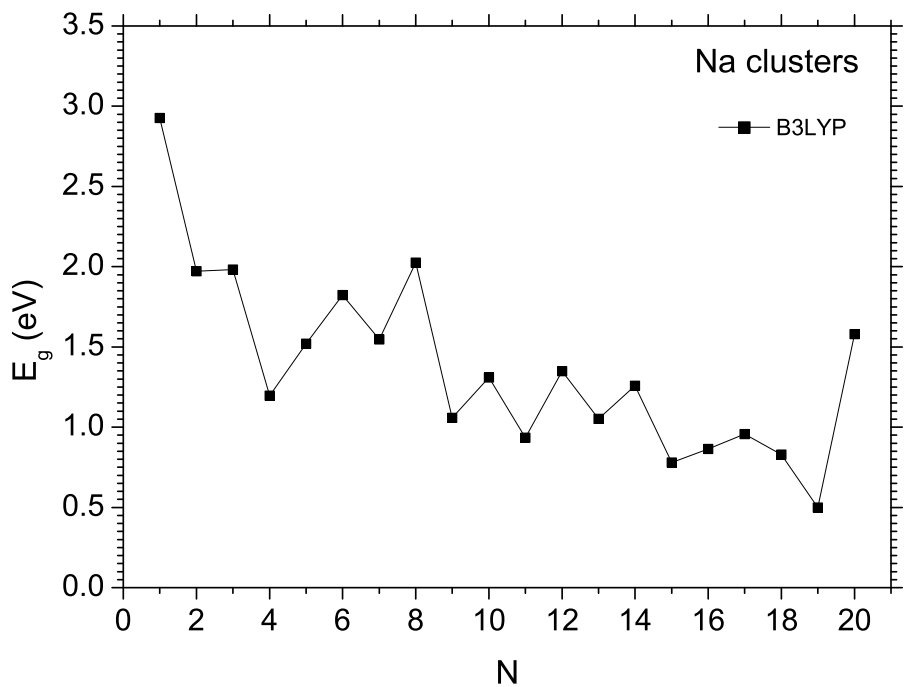}
\caption{Gap between the highest occupied and the lowest unoccupied eigenstates
for the $Mg$ and $Na$ clusters as a function of cluster size.
Circles and squares represent the HOMO-LUMO gap calculated
by the $B3PW91$ and the $B3LYP$ methods respectively.
Stars show the result of the work
by Akola {\it et al} \cite{Akola01}.}
\label{homo_lumo}
\end{figure}

Figure \ref{homo_lumo} shows the gap $E_{g}$ between the highest occupied and 
the lowest unoccupied molecular orbitals (HOMO-LUMO gap) for the $Mg$-clusters 
as a function of cluster size. For the sake of comparison, we have also calculated the 
HOMO-LUMO gap for the sodium clusters and present it in  figure \ref{homo_lumo}.
Calculations have been performed using the $B3PW91$ and $B3PLYP$ methods. 
The geometries of neutral sodium clusters
have been taken from \cite{StructNa}. 
For the small magnesium clusters with $N \le 13$, we
compare our results with those presented 
in Ref. \cite{Akola01}.

The size dependence of $E_{g}$ for neutral sodium clusters has an 
oscillatory behavior with local maxima at $N=$ 6, 8, 10, 14 and 20. 
These  maxima correspond to the 
electronic shell closures in a full accordance  with the deformed jellium model.
The local maximum in the size dependence of $E_{g}$ at $N = 12$
and the shift of the local maximum from $N = 18$ to $N = 17$ are the consequences of
triaxial deformations \cite{StructNa}. 
Thus, the triaxial deformation leads to the splitting of the fourfold degenerated 
highest occupied orbital on two twofold degenerated orbitals. 
As a result of that the additional shell closure at the $N_{el}=12$ appears. 
  
For $Mg$-clusters, the evolution of the HOMO-LUMO gap with the growth cluster size
differs from that for $Na$-clusters.
The gap $E_{g}$ calculated for magnesium clusters shows the oscillatory 
behavior accompanied by the gradual decrease in the absolute value. 
Maxima in this dependence at $N=$ 4, 10 and 20
correspond to the magic numbers of the spherical jellium model 
($N_{el}=$ 8, 20 and 40 respectively). The similar feature also does exist 
for $Na$-clusters at  $N=$ 8 and 20. 
Additional variation of $E_{g}$ appears both due the  
subshell closures and the cluster structural rearrangements.

We note that the HOMO-LUMO gap remains  rather large even
for the clusters with $N \ge 15$ possessing elements of the 
{\it hcp} lattice of the bulk $Mg$.
This fact confirms the conclusion on the slow 
and non-monotonous evolution of metallic properties in $Mg$ clusters.

\section{Conclusion}
\label{conclusion}

The optimized geometries and electronic properties of neutral and singly-charged
magnesium  clusters consisting of
up to 21 atoms have been investigated using the $B3PW91$, $B3LYP$ and 
$MP4$ methods accounting for all electron in the system. 
The detailed comparison of the results
of the phenomenological
$B3PW91$ and $B3LYP$ density-functional methods with the results of the 
systematic {\it ab initio} post-Hartree-Fock
many-body theory leads us to the conclusion
that the $B3PW91$ method is more reliable for
$Mg$-cluster simulations than the $B3LYP$ one. 

We have investigated the size  evolution of the $Mg$-clusters geometry. It has been
shown that starting from  $Mg_{15}$ the hexagonal ring structure
determines the cluster growth, which is the basic element of the 
{\it hcp} lattice for the bulk magnesium.

We have investigated the electronic properties of magnesium clusters.
It has been shown that the electronic shell effects and jellium-like behavior 
clearly manifest themselves in the formation of geometrical 
properties, however, the shell effects do not determine the geometry 
of $Mg$ clusters completely. We have demonstrated that due to 
the interplay of the jellium and  the {\it hcp} lattice factors  
the electronic properties of magnesium clusters 
possess irregularities which can not be explained within the simple jellium model.
It has been shown that the metallic evolution of magnesium clusters is slow and 
non-monotonous  process.

The results of this work can be extended in various directions.
One can use the similar methods to study structure and
properties of various types of clusters. It is interesting to
extend calculations towards larger cluster sizes and to perform
more advanced comparison of model and {\it ab initio} approaches.
A lot of novel problems arise,
when considering collisions and electron excitations in the
clusters with the optimized geometries \cite{AVSol}. These and many more 
other problems on atomic cluster physics can be tackled
with the use of methods considered in our work.

\begin{acknowledgments}
The authors acknowledge support from
the Alexander von Humboldt Foundation and
the Russian Academy of Sciences (Grant 44).
\end{acknowledgments}

{\newpage
\appendix*
\section{}
\label{appendix}

\begingroup
\squeezetable
\begin{table*}[hp]
\caption{Total energies and the point symmetry groups for a variety of isomers of  
neutral magnesium clusters. 
Calculations have been done by the $B3LYP$ method.}
\label{tab:B3LYP}

\begin{ruledtabular}
\begin{tabular}{ccccccccc}
 
     \multicolumn{1}{c}{N } &
     \multicolumn{1}{c}{Point group} &
     \multicolumn{1}{c}{Energy (a.u.)} &
     \multicolumn{1}{c}{N } &
     \multicolumn{1}{c}{Point group} &
     \multicolumn{1}{c}{Energy (a.u.)}&
     \multicolumn{1}{c}{N } &
     \multicolumn{1}{c}{Point group} &
     \multicolumn{1}{c}{Energy (a.u.)}\\

\hline 
1 &                & -200.0931   &  9 & $D_{\infty h}$ & -1800.8472  & 16 & $d.C_{s}$      & -3201.6766\\                                    
2 & $D_{\infty h}$ & -400.1868   & 10 & $C_{3v}$       & -2001.0336  &    & $c.C_{s}$      & -3201.6713\\                                    
3 & $D_{3h}$       & -600.2840   &    & $C_{4v}$       & -2001.0239  &    & $b.C_{1}$      & -3201.6673\\                                    
  & $D_{\infty h}$ & -600.2807   &    & $T_{d}$        & -2001.0225  &    & $a.C_{1}$      & -3201.6646\\                                    
4 & $T_{d}$        & -800.3938   &    & $C_{1}$        & -2001.0120  &    & $T_{d}$        & -3201.6594\\
  & $D_{2h}$       & -800.3792   &    & $D_{3h}$       & -2001.0114  &    & $a.C_{s}$      & -3201.6579\\
  & $D_{\infty h}$ & -800.3750   &    & $C_{2v}$       & -2001.0098  &    & $b.C_{s}$      & -3201.6572\\
  & $D_{4h}$       & -800.3748   &    & $D_{4d}$       & -2001.0078  &    & $D_{\infty h}$ & -3201.5082\\
5 & $D_{3h}$       & -1000.4907  &    & $D_{\infty h}$ & -2000.9417  & 17 & $D_{4d}$       & -3401.7953 \\
  & $C_{4v}$       & -1000.4787  &    & $D_{10h}$      & -2000.9322  &    & $C_{s}$        & -3401.7861\\
  & $T_{d}$        & -1000.4768  & 11 & $D_{3h}$       & -2201.1348  &    & $D_{3h}$       & -3401.7052\\
  & $C_{2v}$       & -1000.4766  &    & $D_{11h}$      & -2201.0430  &    & $D_{\infty h}$ & -3401.6026\\
  & $D_{5h}$       & -1000.4751  &    & $D_{\infty h}$ & -2201.0362  & 18 & $b.C_{2v}$     & -3601.9095\\
  & $D_{2D}$       & -1000.4743  & 12 & $a.C_{s}$      & -2401.2366  &    & $c.C_{s}$      & -3601.9026\\
  & $D_{2h}$       & -1000.4738  &    & $b.C_{s}$      & -2401.2304  &    & $C_{2}$        & -3601.8764\\
  & $D_{\infty h}$ & -1000.4694  &    & $C_{2v}$       & -2401.2178  &    & $b.C_{s}$      & -3601.8752\\
6 & $C_{2v}$       & -1200.5898  &    & $D_{6h}$       & -2401.1313  &    & $a.C_{s}$      & -3601.8716\\
  & $D_{2h}$       & -1200.5897  &    & $D_{\infty h}$ & -2401.1308  &    & $a.C_{2v}$     & -3601.8711\\                                    
  & $D_{4h}$       & -1200.5851  &    & $D_{12h}$      & -2401.1278  &    & $D_{5h}$       & -3601.8608\\
  & $C_{5v}$       & -1200.5815  & 13 & $b.C_{1}$      & -2601.3438  &    & $D_{6h}$       & -3601.7022\\
  & $D_{3h}$       & -1200.5765  &    & $C_{S}$        & -2601.3308  &    & $D_{\infty h}$ & -3601.6970\\
  & $O_{h}$        & -1200.5720  &    & $a.C_{1}$      & -2601.3302  & 19 & $C_{2v}$       & -3802.0292\\
  & $D_{6h}$       & -1200.5639  &    & $C_{3v}$       & -2601.3221  &    & $C_{3v}$       & -3801.9867\\                                    
  & $D_{\infty h}$ & -1200.5638  &    & $I_{h}$        & -2601.2904  &    & $D_{5h}$       & -3801.9706\\
7 & $D_{5h}$       & -1400.6933  &    & $D_{6h}$       & -2601.2401  &    & $D_{6h}$       & -3801.8449\\
  & $C_{3}$        & -1400.6908  &    & $D_{\infty h}$ & -2601.2253  &    & $D_{\infty h}$ & -3801.7925\\               
  & $C_{3v}$       & -1400.6854  & 14 & $C_{1}$        & -2801.4485  & 20 & $C_{s}$        & -4002.1503\\ 
  & $D_{7h}$       & -1400.6645  &    & $b.C_{3v}$     & -2801.4397  &    & $C_{1}$        & -4002.1444\\               
  & $D_{\infty h}$ & -1400.6583  &    & $C_{s}$        & -2801.4407  &    & $d.C_{2v}$     & -4002.1392\\                                    
8 & $a.C_{2v}$     & -1600.7999  &    & $C_{2v}$       & -2801.4328  &    & $c.C_{2v}$     & -4002.1263\\               
  & $C_{s}$        & -1600.7976  &    & $O_{h}$        & -2801.4115  &    & $b.C_{2v}$     & -4002.1212\\               
  & $b.C_{2v}$     & -1600.7948  &    & $a.C_{3v}$     & -2801.4134  &    & $T_{d}$        & -4002.1026\\               
  & $T_{d}$        & -1600.7854  &    & $D_{\infty h}$ & -2801.3194  &    & $a.C_{2v}$     & -4002.0990\\               
  & $D_{\infty h}$ & -1600.7527  & 15 & $D_{3h}$       & -3001.5692  &    & $D_{6h}$       & -4001.9764\\               
9 & $D_{3h}$       & -1800.9162  &    & $C_{s}$        & -3001.5627  &    & $D_{\infty h}$ & -4001.8870\\
  & $C_{3v}$       & -1800.9064  &    & $D_{6h}$       & -3001.4594  & 21 & $C_{1}$        & -4202.2460\\  
  & $D_{9h}$       & -1800.8540  &    & $D_{\infty h}$ & -3001.4138  &    & $C_{2v}$       & -4202.2255\\

\end{tabular}
\end{ruledtabular}
\end{table*}
\endgroup
}

\begingroup
\squeezetable
\begin{table*}[hp]
\caption{The same as table \ref{tab:B3LYP} but for the $P3PW91$ method.}
\label{tab:B3PW91}

\begin{ruledtabular}

\begin{tabular}{ccccccccc}
 
     \multicolumn{1}{c}{N } &
     \multicolumn{1}{c}{Point group} &
     \multicolumn{1}{c}{Energy (a.u.)} &
     \multicolumn{1}{c}{N } &
     \multicolumn{1}{c}{Point group} &
     \multicolumn{1}{c}{Energy (a.u.)}\\

\hline 

1 &                & -200.0379   & 14 & $C_{1}$        & -2800.8170\\                                    
2 & $D_{\infty h}$ & -400.0788   &    & $D_{\infty h}$ & -2800.6007\\                                    
3 & $D_{3h}$       & -600.1281   &    & $D_{14 h}$     & -2800.5997\\                                    
  & $D_{\infty h}$ & -600.1207   & 15 & $D_{3h}$       & -3000.9045\\                                    
4 & $T_{d}$        & -800.1962   &    & $C_{s}$        & -3000.8914\\
  & $D_{4h}$       & -800.1638   &    & $D_{15 h}$     & -3000.6786\\
  & $D_{\infty h}$ & -800.1633   &    & $D_{\infty h}$ & -3000.6453\\
5 & $D_{3h}$       & -1000.2450  & 16 & $d.C_{s}$      & -3200.9666\\
  & $D_{5h}$       & -1000.2193  &    & $c.C_{s}$      & -3200.9560\\
  & $D_{\infty h}$ & -1000.2063  &    & $D_{\infty h}$ & -3200.6902\\ 
6 & $C_{2v}$       & -1200.2970  &    & $D_{16 h}$     & -3200.6880\\
  & $D_{2h}$       & -1200.2966  & 17 & $D_{4d}$       & -3401.0612\\
  & $C_{5v}$       & -1200.2865  &    & $C_{s}$        & -3401.0428\\
  & $D_{6h}$       & -1200.2495  &    & $D_{17 h}$     & -3400.7719\\
  & $D_{\infty h}$ & -1200.2495  &    & $D_{\infty h}$ & -3400.7353\\ 
7 & $D_{5h}$       & -1400.3604  & 18 & $b.C_{2v}$     & -3601.1279\\
  & $D_{7h}$       & -1400.3091  &    & $c.C_{s}$      & -3601.1221\\
  & $D_{\infty h}$ & -1400.2928  &    & $b.C_{s}$      & -3601.0756\\                                    
8 & $C_{s}$        & -1600.4205  &    & $D_{\infty h}$ & -3600.7806\\
  & $a.C_{2v}$     & -1600.4193  &    & $D_{18 h}$     & -3600.7766\\
  & $T_{d}$        & -1600.4005  & 19 & $C_{2v}$       & -3801.2093\\
  & $D_{\infty h}$ & -1600.3364  &    & $D_{19 h}$     & -3800.8653\\ 
  & $D_{8h}$       & -1600.3335  &    & $D_{\infty h}$ & -3800.8260\\                                    
9 & $D_{3h}$       & -1800.5018  & 20 & $C_{s}$        & -4001.2920\\
  & $D_{9h}$       & -1800.4000  &    & $C_{1}$        & -4001.2801\\
  & $D_{\infty h}$ & -1800.3800  &    & $b.C_{2v}$     & -4001.2539\\               
10& $C_{3v}$       & -2000.5786  &    & $D_{20 h}$     & -4000.9111\\ 
  & $T_{d}$        & -2000.5590  &    & $D_{\infty h}$ & -4000.8716\\               
  & $D_{10h}$      & -2000.4367  & 21 & $C_{1}$        & -4201.3398\\                                    
  & $D_{\infty h}$ & -2000.4238  &    & $C_{2v}$       & -4201.3201\\               
11& $D_{3h}$       & -2200.6322  &    &                &           \\
  & $D_{11h}$      & -2200.4923  &    &                &           \\                                                                 
  & $D_{\infty h}$ & -2200.4678  &    &                &           \\                                                                 
12& $a.C_{s}$      & -2400.6891  &    &                &           \\
  & $D_{12h}$      & -2400.5321  &    &                &           \\
  & $D_{\infty h}$ & -2400.5119  &    &                &           \\
13& $b.C_{1}$      & -2600.7561  &    &                &           \\
  & $D_{13h}$      & -2600.5853  &    &                &           \\
  & $D_{\infty h}$ & -2600.5562  &    &                &           \\                                                                   
                                                                   
\end{tabular}                                                      
\end{ruledtabular}                                                 
\end{table*}

\end{document}